\documentclass[%
reprint,
amsmath,
amssymb,
jmp,
aps,
]{revtex4-1}
\usepackage{graphicx}
\usepackage{bm}
\pdfoutput=1
\begin{document}


\title{Exploiting short-term memory in soft body dynamics as a computational resource}
\author{Kohei Nakajima,$^{1,2*}$ Tao Li,$^{3}$ Helmut Hauser,$^{3}$ Rolf Pfeifer$^{3**}$}
\email{jc\_mc\_datsu@yahoo.co.jp}
\affiliation{$^{1}$The Hakubi Center for Advanced Research, Kyoto University, 606-8501 Kyoto, Japan}
\affiliation{$^{2}$Nonequilibrium Physics and Theoretical Neuroscience, Department of Applied Analysis and Complex Dynamical Systems, Graduate School of Informatics, Kyoto University, 606-8501 Kyoto, Japan}
\affiliation{$^{2}$Department of Informatics, University of Zurich, Andreasstrasse 15, 8050 Zurich, Switzerland.}
\begin{abstract}
Soft materials are not only highly deformable but they also possess rich and diverse body dynamics. 
Soft body dynamics exhibit a variety of properties, including nonlinearity, elasticity, and potentially infinitely many degrees of freedom. 
Here we demonstrate that such soft body dynamics can be employed to conduct certain types of computation. 
Using body dynamics generated from a soft silicone arm, we show that they can be exploited to emulate functions that require memory and to embed robust closed-loop control into the arm. 
Our results suggest that soft body dynamics have a short-term memory and can serve as a computational resource. 
This finding paves the way toward exploiting passive body dynamics for control of a large class of underactuated systems.
\end{abstract}

\maketitle

In recent years soft materials have been increasingly used to incorporate flexible elements into robots' bodies. 
The resulting machines, called {\it soft robots}, have significant advantages over traditional articulated robots due to deformable morphology and safety in interaction \cite{Rolf3}. 
They can adapt their morphology to unstructured environments, and carry and touch fragile objects without causing damage, which makes them applicable for rescue and human interactions, in particular care for the elderly, prosthetics, and wearables \cite{Trivedi,Cecilia}. 
In addition, they can generate diverse behaviors with simple types of actuation by partially outsourcing control to the morphological and material properties of their soft bodies \cite{Whitesides1}, which is made possible by the tight coupling between control, body, and environment \cite{Rolf1,Rolf2}. 
In this paper, we build on these perspectives and add a novel advantage of soft bodies, demonstrating that they can be exploited as computational resources.

One of the major differences between rigid and soft bodies can be found in their body dynamics. 
Soft body dynamics usually exhibit a variety of properties, including nonlinearity, elasticity, and potentially infinitely many degrees of freedom, 
which are difficult to reduce to lower dimensionality.
In particular, their degrees of freedom are often larger than a number of actuators, which leads to a typical underactuated system \cite{underactuated}, and this makes the soft body difficult to control with conventional frameworks.
Here we demonstrate that these properties can in fact be highly beneficial in that they can be employed for computation. 
Our approach is based on a machine learning technique called {\it reservoir computing}, which has a particular focus on real-time computing of time-varying input that provides an alternative to computational frameworks based on Turing machines \cite{Jaeger1,Jaeger2,Maass,Reservoir}. 
By driving a high-dimensional dynamical system, typically referred to as the {\it reservoir}, with a low-dimensional input stream, transient dynamics are generated that operate as a type of temporal and finite kernel that facilitate the separation of input states \cite{Maass,Transient}. 
If the dynamics involve enough nonlinearity and memory, emulating complex nonlinear dynamical systems only requires adding a linear, static readout from the high-dimensional state space of the reservoir.
A number of different implementations for reservoirs have been proposed: for example, abstract dynamical systems for echo state networks (ESN) \cite{Jaeger1,Jaeger2}, or models of neurons for liquid state machines \cite{Maass}. 
Implementations even include using the surface of water in a laminar state \cite{Bucket}. 
Lately, it has been demonstrated that nonlinear mass spring systems have the potential to serve as reservoirs as well \cite{Helmut1,Helmut2}, and this has been applied in a number of ways (see, e.g., \cite{Kohei,Kohei_icra,Ken}).

In this study, we establish a simple but powerful physical platform with a soft silicone arm and demonstrate, through a number of experiments, that the soft body dynamics can be used as a reservoir. 
In particular, we focus on the property of short-term memory \cite{Jaeger3,White,Surya}, which is the ability to store information about recent input sequences in the transient dynamics of the reservoir. 
In neuroscience, this property has drawn attention as a mechanism to perform real-time computations on sensory input streams \cite{Buonomano,Hausler}, which is a prerequisite for cognitive phenomena, such as planning and decision making. 
We show that short-term memory also exists in the body dynamics of a soft silicone arm and, in particular, that it can be exploited to control the arm's motions robustly in a closed-loop manner.
In other words, the seemingly undesirable properties of soft body dynamics are no longer drawbacks for control but constitute core aspects of the system's functionality.

\section*{Results}
\subsection*{A soft silicone arm as a computational resource}
\begin{figure*}[htbp]
\centerline{\includegraphics[width=3.0in,bb=98 50 523 773]{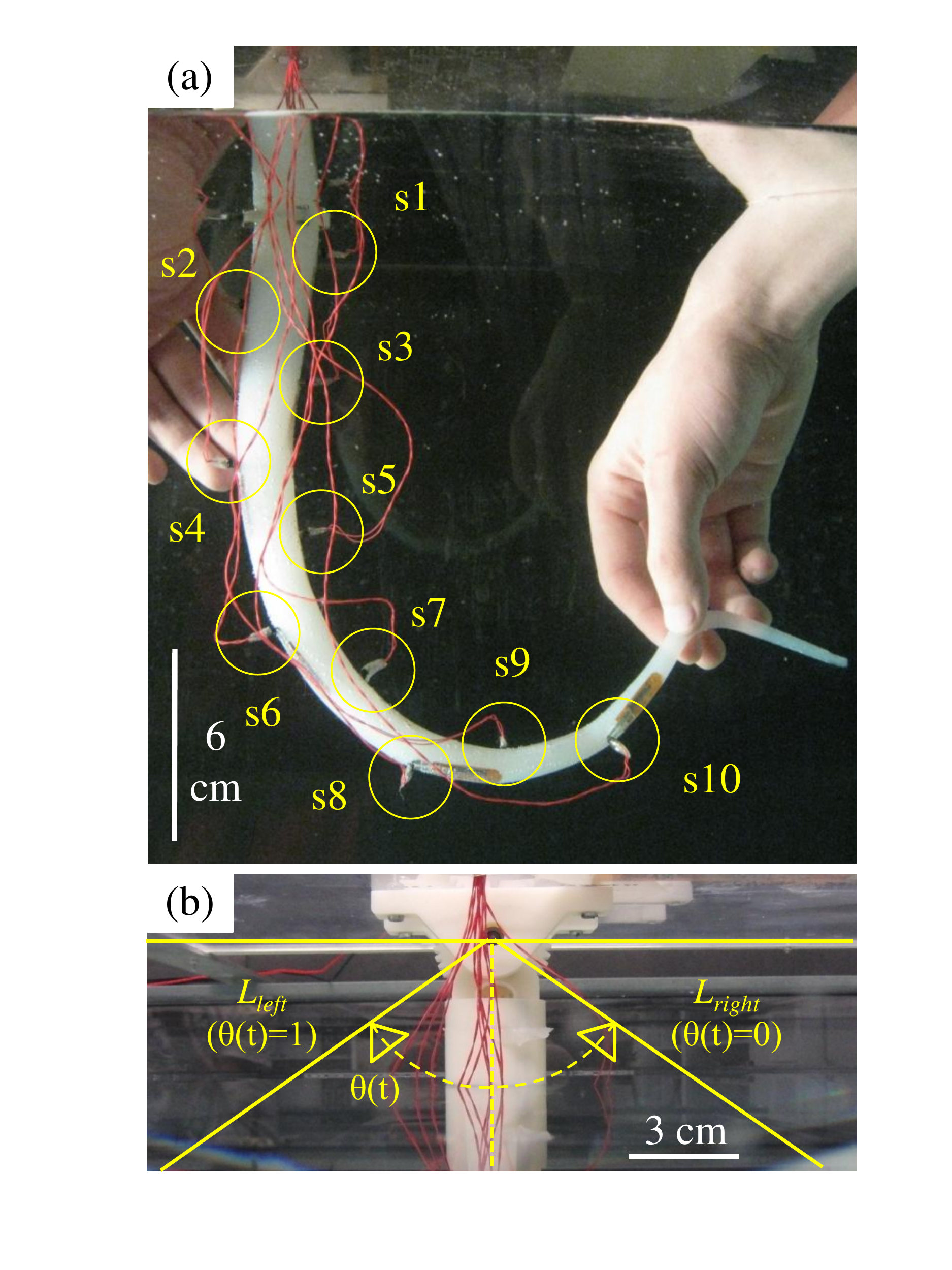}} 
\caption{Platform setup for a soft silicone arm. (a) A soft silicone arm, which contains 10 bend sensors, is immersed underwater. Sensors are connected to a sensory board by the red wires. The wires are set as carefully as possible so as not to affect the arm motion. (b) Motor commands take binary states. When these commands are set to $0$ ($1$), the base of the arm rotates to the right (left) hand side toward $L_{right}$ ($L_{left}$). See the main text for details.} \label{platform}
\end{figure*}

There have been several soft silicone arms proposed in the literature, which are inspired by the octopus (see, e.g., \cite{Matteo,Machelo,Whitesides2}).
In this paper, we use a soft silicone arm, which has a similar material characteristic to the one proposed in \cite{Matteo}.
The platform consists of a soft silicone arm, its sensing and actuation systems, data processing via a PC, and a water tank containing fresh water as an underwater environment (Fig.\ref{platform}).
By rotating the base of the arm and generating body dynamics induced by the interaction between the underwater environment and the soft silicone material, we aim to show that the sensory time series that are reflected in the body dynamics can be exploited as part of a computational device. 
The unit of timestep $t$ used in this study is a sensing and actuation loop of the PC (this is approximately 0.03 $s$ in physical time). 
Throughout this study, we observe the behavior of the system from one side of the tank and use terminology, such as ``left'' or ``right,'' with respect to this point of view.

\begin{figure*}[htbp]
\centerline{\includegraphics[width=5.5in, bb=33 375 563 718]{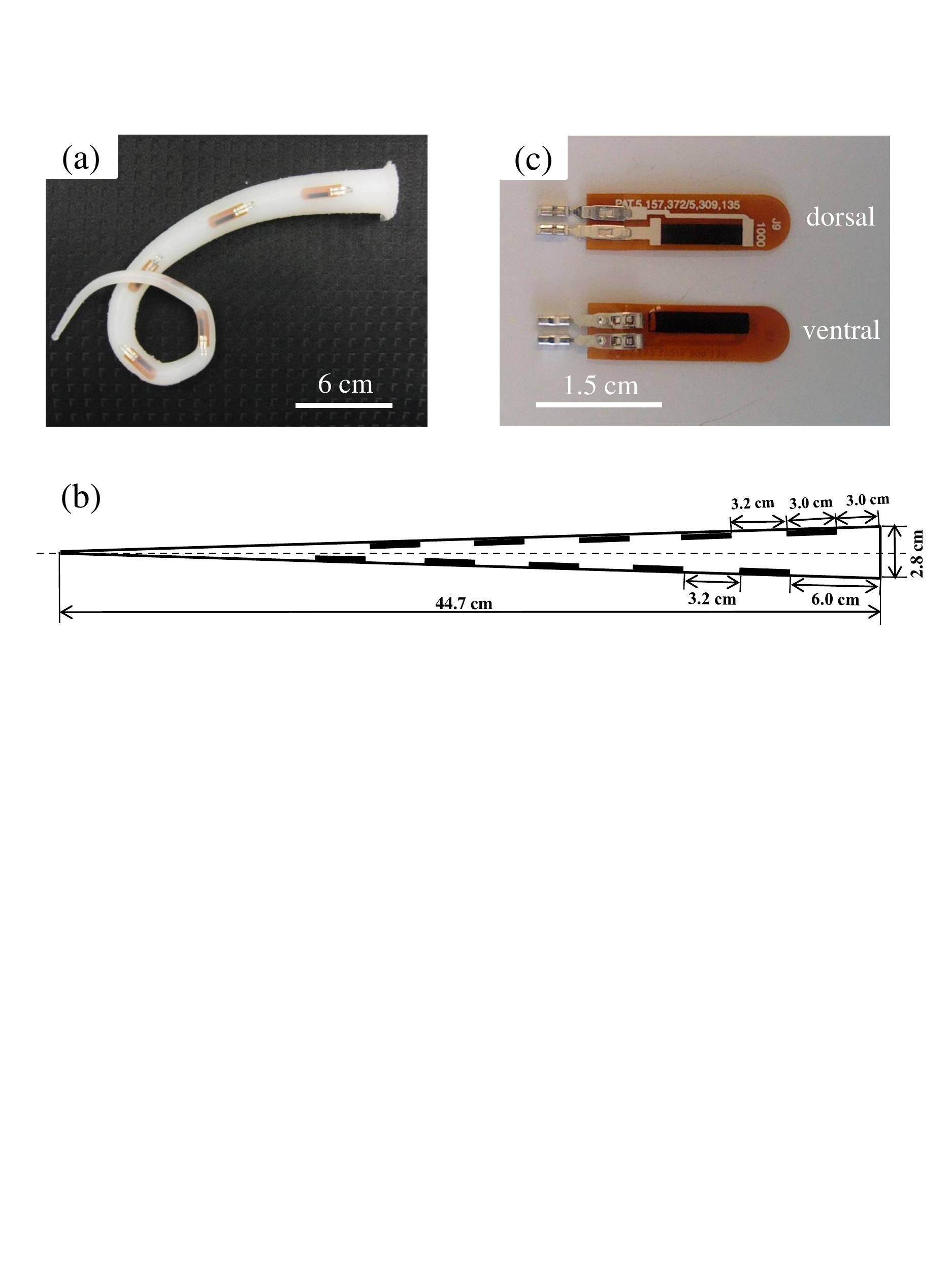}} 
\caption{Soft silicone arm and bend sensors. (a) A soft silicone arm with 10 embedded bend sensors. (b) Schematics showing the alignment of the bend sensors in the arm. The sensors are aligned parallel to the arm surface with an equal distance of 3.2 cm between them. There is a thin layer of silicone (about 0.1 cm) covering the sensors. (c) The bend sensors. Both dorsal and ventral sides are shown.} \label{arm}
\end{figure*}

The arm embeds 10 bend sensors within the silicone material (Fig.\ref{platform}(a) and Fig.\ref{arm}).
A bend sensor gives a base value when it is straight.
If it bends in the ventral side, the sensor value is smaller, and if it bends in the dorsal side, the value is larger; the change in value reflects the degree of bend in each case.
The sensors are embedded near the surface of the arm, with their ventral sides directed outward. 
We numbered these sensors from the base toward the tip as $s1$ through $s10$.
The sensors are embedded alternately, with odd-numbered sensors on the right side of the arm and even-numbered sensors on the left (Figs. \ref{platform}(a) and \ref{arm}).
The base of the arm can rotate left and right through the actuation of a servo motor. 
The motor commands sent from the PC are binary values, $M = \{0, 1\}$.
If the command is $0$ or $1$, the motor is controlled to move from its current position toward the maximum right position ($L_{right}$) or the maximum left position ($L_{left}$), respectively (Fig.\ref{platform}(b)). 
The actual servo motor positions are also sent to the PC to monitor the current position of the base rotation $\theta (t)$.
The positions $L_{right}$ and $L_{left}$ were heuristically determined to avoid damaging the motor components. 
The values for $| L_{right} |$ ($= |L_{left} |$) are about 46.4 degrees by setting the origin of the rotation angle (0 degrees) when the arm is aligned vertically to the water surface.
Throughout this study, $\theta (t)$ is linearly normalized to be in the range from 0 to 1.
Note that the motor command does not always take the roller position to $L_{right}$ or $L_{left}$; rather, it decides the motor movement direction for each timestep. 
In addition, if the command is $0$ or $1$, when the current position is $L_{right}$ or $L_{left}$, respectively, then the position will stay unchanged.

\begin{figure*}[htbp]
\centerline{\includegraphics[width=4.0in,bb=55 200 525 721]{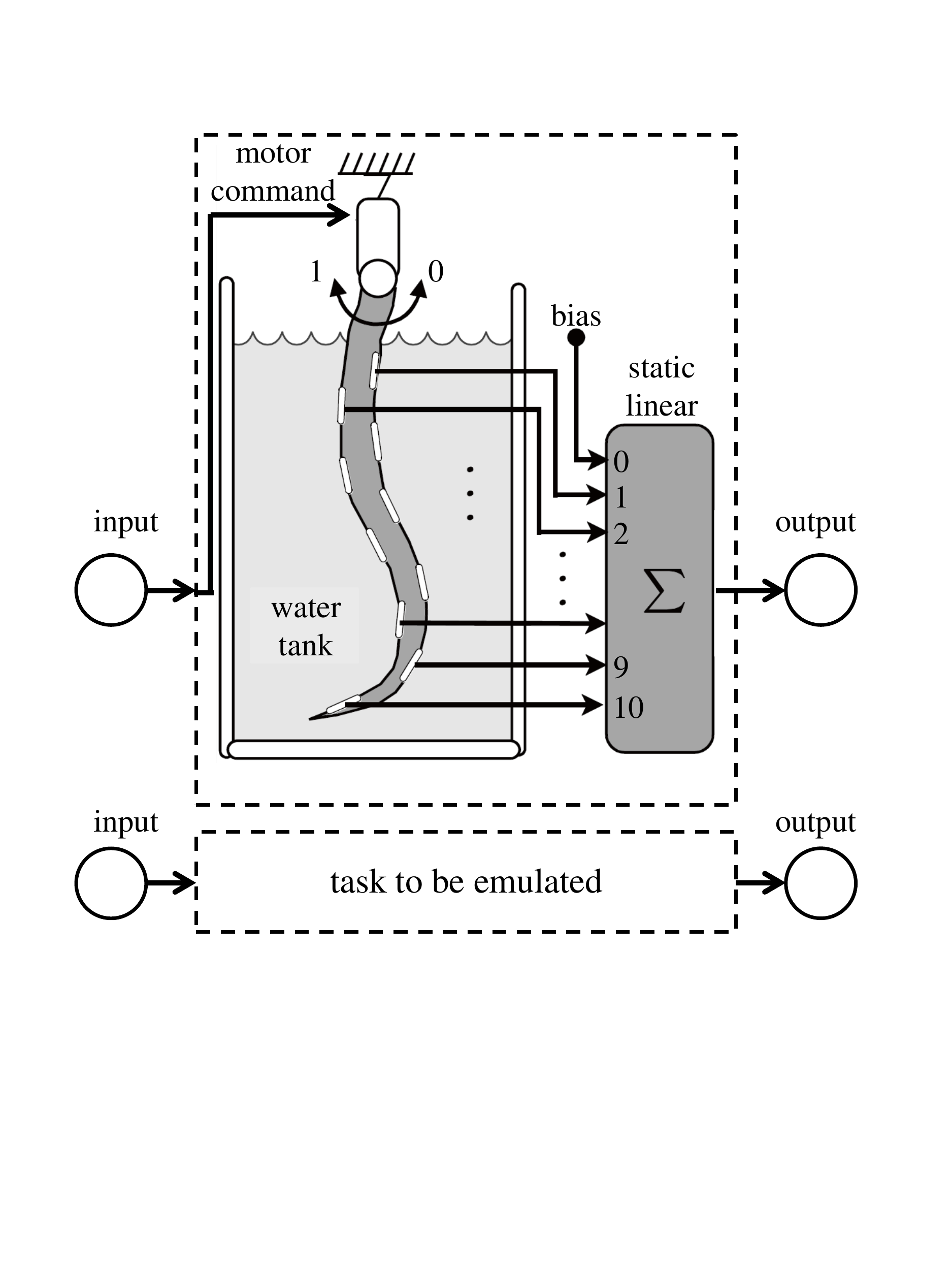}} 
\caption{Schematics showing the information processing scheme using the arm. Input is provided to the motor command to generate arm motion, and the embedded bend sensors reflect the resulting body dynamics. By using the detected sensory time series, the binary state output is generated by thresholding the weighted sum of the sensory values. See the main text for details.} \label{scheme}
\end{figure*}

To exploit the soft silicone arm as a computational resource, we need to determine how to provide inputs $I(t)$ to the system and how to generate corresponding outputs $O(t)$. 
In this paper, we provide the input to the motor command, $m(t) \in M$, 
and the output is generated by linearly combining all sensory time series $s_{i} (t)$ ($i=1, 2, ..., 10$) with a weighted sum using the weights $w_{i}$ ($i=1, 2, ..., 10$) (Fig.\ref{scheme}).
In addition, a bias is added, which is expressed as $b = w_{0} s_{0} (t)$, where $s_{0} (t)$ is a constant value set to 1.
As a result, we have 11 pairs of weights and corresponding sensory time series $(w_{i}, s_{i} (t))$ ($i=0, 1, ..., 10$) in our system.
Our system output takes a binary state, $O(t) \in \{ 0, 1 \}$, which is obtained by thresholding the weighted sum of the sensory values (see Method section for details). 
To emulate a desired function with our system, we first apply the inputs to the system, which then generate the arm motions, and we collect the corresponding sensory time series.
Together with the target outputs, we have a training data set for supervised learning.
The linear readout weights are then optimized with simple logistic regression with respect to minimizing the error between the system output and the target output.
The performance of the system output is evaluated by comparing with the target output for a new experimental trial (see Method section for details on the training procedures and the logistic regression).

We used three tasks to evaluate the computational power of our soft silicone arm with the focus on the property of short-term memory.
Unlike a conventional computer, our system does not contain explicit memory storage; instead, the memory is expected to be implicitly included in the transient dynamics of the soft body.
By assigning a task to the system that requires memory to be carried out and by evaluating its performance, we can characterize its memory capacity.

Our first task is to construct a timer exploiting the soft body dynamics.
Triggered by a cue sent at certain timesteps, the arm starts to move from $L_{right}$ to $L_{left}$.
The system should output a pulse of predefined length by exploiting the body dynamics.
To perform this task, the system has to be able to ``recognize'' the duration of time that has passed since the cue was launched. 
This clearly requires memory.
By increasing the desired pulse response, we systematically investigate the limits of the physical system to represent memory in its transient body dynamics.

The second task is to perform a closed-loop control exploiting the soft body dynamics.
With a periodic square wave function, which switches its motor command from 0 to 1 and from 1 to 0 with a fixed period as a target function, 
we aim to evaluate the maximal length for the period of the square function that can be embedded in the system.
In this task, the system should ``recognize'' how much time has passed since the motor command switched from 0 to 1 (or from 1 to 0), and it should decide when to switch the motor command to the next position.
Again, this task requires memory.  
Furthermore, this task also evaluates whether the soft body dynamics can be exploited as a computational resource to control the arm's own motion.
This is especially interesting as typically the complex dynamics of a soft body are the main obstacles to apply a classic control theoretic approach. 
Remarkably, in our proposed context, this property is beneficial because it can be exploited as a computational resource.

The third task is an emulation task of functions that require memory.
A random binary input sequence is provided to the system, and by exploiting the generated soft body dynamics, the system should emulate two functions simultaneously; the first one is a function that reproduces past inputs with a given delay 
and the second one is the N-bit parity checker.
Emulations of these functions are commonly used as benchmark tasks to characterize the computational power of the system, and again, both functions require memory.
In particular, these functions should be emulated using the same soft body dynamics at the same time, which points to another remarkable property of the approach (typically referred to as {\it multitasking} \cite{Helmut1}).

In all three tasks, we are only adjusting the linear readout weights, which are fixed after learning, i.e., no memory is present in the readout.
Hence, we can confirm that the required memory is purely due to the property of the soft silicone arm. 
Unlike conventional computational units (e.g., artificial neural networks), our proposed setup has a constraint due to the specifications of the mechanical structure of the system because inputs are transformed to the mechanical realm.
For example, a drastic and frequent switching of the motor command can result in motor overheat and a total stop.
We defined the presented tasks to evaluate the memory capacity of our system by taking these physical constraints into consideration.
Accordingly, the input/output (I/O) setting in our system slightly differs in each task (see Method section for detailed information on the I/O setting for each task).

\subsection*{Dynamic property of the silicone arm}
\begin{figure*}[htbp]
\centerline{\includegraphics[width=6.5in,bb=18 400 577 774]{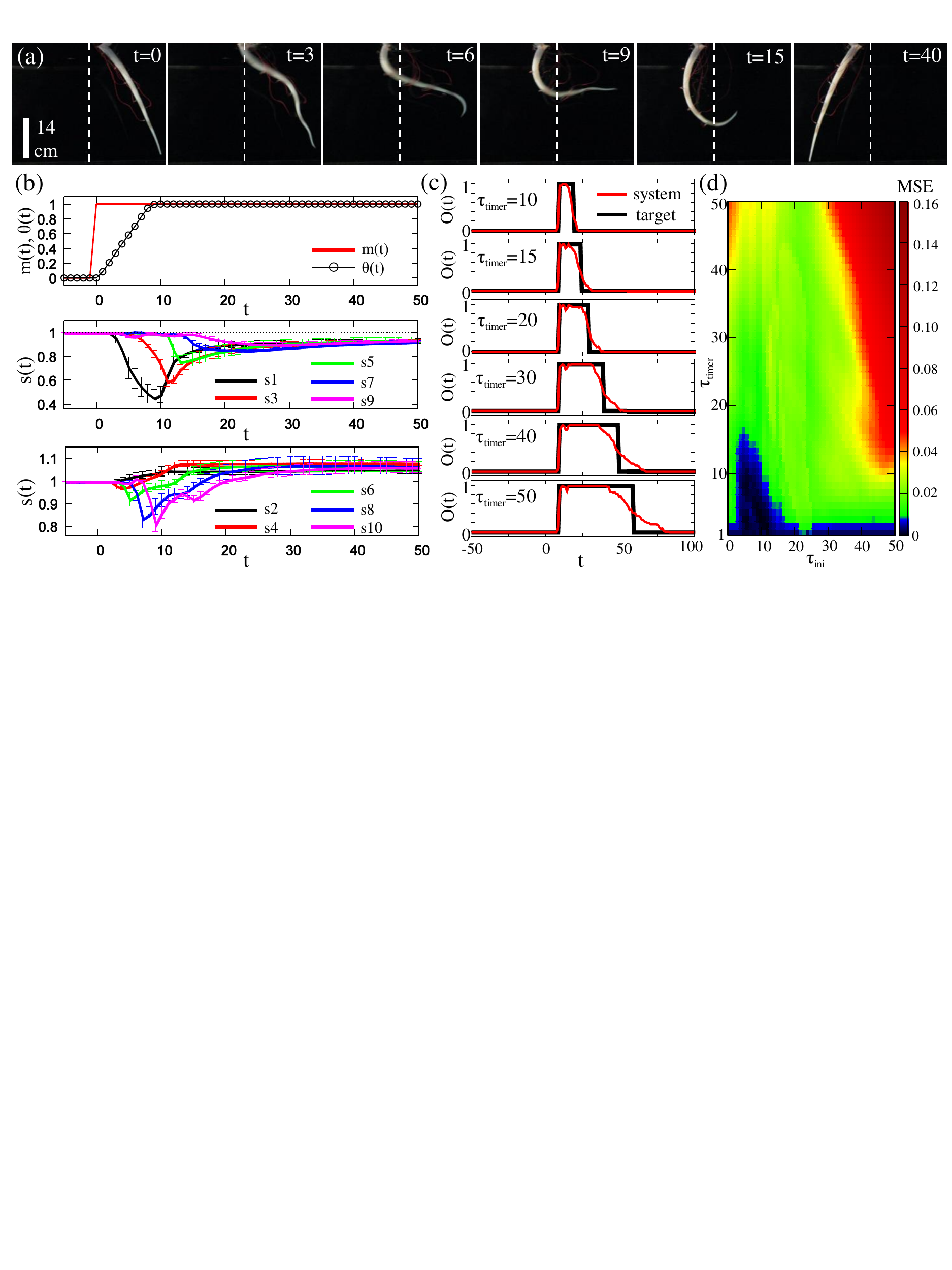}} 
\caption{Sensory response during the arm motion and performance for the timer task. (a) Snapshots showing a typical arm motion when the motor command is switched from 0 to 1 at $t=0$ (i.e., movement from $L_{right}$ to $L_{left}$). (b) Plots showing the dynamics of the motor command $m(t)$ and the normalized base angle $\theta (t)$ (the upper plot) and the corresponding sensory time series $s(t)$ (the lower two plots). The middle and lower plots show the average sensory response curves for the odd- and even-numbered sensors, respectively. For each sensor, the sensory values are linearly scaled to make the sensory values to 1 when the arm is in $L_{right}$ and averaged over 50 trials. The error bars show the standard deviations. (c) The plots show the average system outputs for each $\tau_{timer}$ ($\tau_{timer} = 10, 15, 20, 30, 40,$ and $50$) when $\tau_{ini}$ is fixed to 9. The black lines show the target output and the red lines show the averaged system outputs over 25 trials for each condition. Note that the averaged system outputs can take values in the range of [0, 1]. (d) The plot shows the average MSE over 25 trials with respect to each $\tau_{timer}$ and $\tau_{ini}$ varied from 1 to 50 and from 0 to 50, respectively.} \label{respnse_curve}
\end{figure*}

We here present the basic property of our arm motion and the step response.
Figure \ref{respnse_curve}(a) shows a typical arm motion when the motor command is switched from $0$ to $1$. 
The arm is initially set to $L_{right}$, and at $t=0$, it starts to move toward $L_{left}$.
The silicone arm shows characteristic body dynamics because of the interaction with the water (see Video S1).
In particular, even when the base reaches the position of $L_{left}$, the entire arm still shows transient dynamics.
The figure clearly shows that because the arm moves from right to left, the right side of the arm bends and the left side of the arm arches according to the water friction.

The dynamic behavior of the arm can be captured by the responses of the sensors (Fig.\ref{respnse_curve}(b), Video S1).
When the motor command switches from $0$ to $1$, $\theta (t)$ takes about 9 timesteps to reach $\theta (t) = 1$, which forms a physical constraint based on the motor and the mechanical structure of our platform (Fig.\ref{respnse_curve}(b), upper plot, Video S1).
When the motor command is switched from $0$ to $1$, all the odd-numbered sensors start to show smaller values than those shown before the motion generation.
They take the local minimum at a different timestep, then gradually approach their resting states (Fig.\ref{respnse_curve}(b), middle plot, Video S1).
Because the arm is passive, the movement of the base rotation propagates from the base toward the tip at a certain velocity. 
For example, $s1$ seems to show a direct reflection of the motor actuation because it is embedded close to the base. 
This effect can be confirmed by checking the local minimum of the sensory response of $s1$ at around timestep 9, which is the same timestep at which the motor rotation stops. 
For even-numbered sensors, although all sensors show larger values than the values before the motion generation, some sensors (e.g., $s6$, $s8$, and $s10$) show a smaller value in some timesteps due to inertia caused by the immediate bend in the left side of the arm (Fig.\ref{respnse_curve}(b), lower plot, Video S1).
This effect also seems to be propagating from the base toward the tip of the arm.
All sensors reach a resting state at around 40 timesteps.
In the resting state $L_{left}$, the odd-numbered sensors show smaller values and the even-numbered sensors show greater values than those shown before motion generation (Fig.\ref{respnse_curve}(b), Video S1).
This phenomenon is the result of gravity; the left side of the arm arches slightly, whereas the right side of the arm bends slightly (Fig.\ref{respnse_curve}(a)). 
When the motor command is switched from $1$ to $0$ with the arm position initially set to $L_{left}$, we can observe a similar behavior with switched roles of the odd- and even-numbered sensors.

\subsection*{Timer task}
Our first task is to emulate the function of a timer exploiting the body dynamics of the arm.
The task has been chosen as it enables us to investigate systematically the memory inherently present in the soft body dynamics. 
One of the characteristic properties of our soft body is its transient dynamics during its motion from one state to another, e.g., moving from right to left.
In this task, the arm is initially set to $L_{right}$ and kept at this position.
Triggered by the input at $t_{start}$, the motor command switches from 0 to 1, when the rotation of the base generates the body dynamics (Fig.\ref{respnse_curve}(a)).
The timer task consists of producing an output pulse starting from $\tau_{ini}$ timesteps after $t_{start}$, which is $\tau_{timer}$ timesteps in length, by exploiting the body dynamics during this transient single motion (see Method section and Fig.\ref{SI_task1} in it for details). 
To perform this task, the system has to have a certain amount of memory. 
In other words, we can evaluate whether the sensory time series that reflects the transient dynamics during the motion from $L_{right}$ to $L_{left}$ contains sufficient information to recognize the duration of time since the trigger event by applying this task. 
A similar task was introduced in \cite{Jaeger1} to demonstrate the existence of short-term memory within an artificial recurrent neural network (e.g., ESN) \cite{Jaeger2,Jaeger3}.
To demonstrate that such a memory can be found and exploited in a real physical system, we applied this task employing the soft silicone arm.
As explained earlier, our system output is generated by thresholding the weighted sum of the sensory values, and the weights are optimized with a simple logistic regression by using a data set collected in the training phase (see Method section for details).
We performed this experiment by varying $\tau_{ini}$ and $\tau_{timer}$ to investigate the relevance of these parameters to the system performance.

Figure \ref{respnse_curve}(c) shows examples of the averaged system outputs for each $\tau_{timer}$ with $\tau_{ini}$ fixed to 9.
As one can see, our system is able to emulate a timer with given duration times $\tau_{timer}$ (see also Video S1). 
Naturally, the performance decreases when increasing the length of $\tau_{timer}$.
This is caused by the gradual fading of memory within the body dynamics after the initiation of motion generation.
This tendency can be found for different settings of $\tau_{ini}$ (Fig.\ref{respnse_curve}(d)).
As can be seen in Fig.\ref{respnse_curve}(d), the error values (mean squared error (MSE)) are especially low when around $\tau_{ini} < 20$ and $\tau_{timer} < 20$, characterizing the amount of memory that can be exploited with the given soft body.
Note that when $\tau_{ini}$ is close to 0, the error values are higher than for other parameters.
This is because when the arm starts to move, the effect of the motor rotation takes some time to propagate due to the softness of the arm (Figs.\ref{respnse_curve}(a), (b)), and if $\tau_{ini}$ is small, it is difficult to distinguish the sensory values from the values when the arm is stopped.

\subsection*{Closed-loop control task}
\begin{figure*}[htbp]
\centerline{\includegraphics[width=3.5in, bb=119 108 463 724]{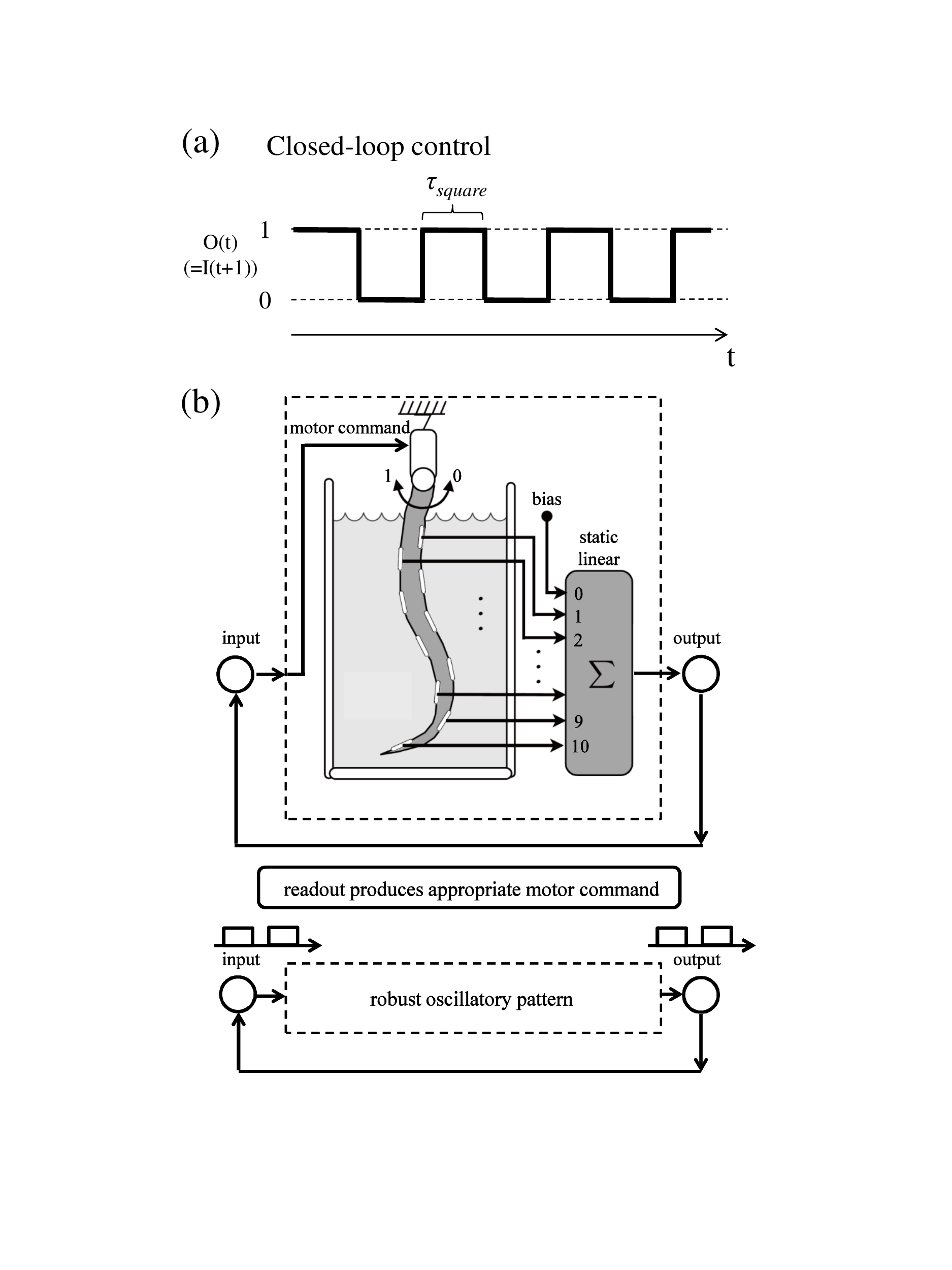}} 
\caption{Experimental procedure for the closed-loop control task. (a) Schematics showing the I/O relation with respect to the temporal axis. The target motor command is a square function with $\tau_{square}$. (b) Schematics showing the generation of the system output in the closed-loop control task. The closed-loop control is realized by feeding back the generated output to the input at the next timestep.} \label{SI_task2}
\end{figure*}

We demonstrated in the previous task that we can use the sensory time series generated by the transient dynamics to construct a timer.
By using the same property, in this second task we aim to realize a closed-loop control of our soft silicone arm.
That is, we aim to demonstrate that the arm's body dynamics can be used to control its own motion.
The target motor command sequence is a square wave in which the amplitude alternates at a steady frequency, between $m(t)=0$ and $1$, with the same duration of timesteps, $\tau_{square}$ (see Method section and Fig.\ref{SI_task2}(a) in it for details).
Similar to the process in the previous task, when the motor command is switched from 0 to 1 (or from 1 to 0), it should recognize the time length of $\tau_{square}$ timesteps and switch the motor command from 1 to 0 (or from 0 to 1).
Thus, it requires memory to fulfill this task.
Recently, similar types of oscillatory motor command have been used to demonstrate the octopus-inspired swimming motion, called {\it sculling}, in a physical platform with an open-loop manner \cite{swimming}.
We aim to emulate this oscillatory wave pattern by using the sensory time series from the soft body and close the loop.
This is realized by feeding back the system output generated by thresholding the weighted sum of the sensory time series as the next motor command to the system (see Method section and Fig.\ref{SI_task2}(b) in it for details).
As with the previous task, we aim to emulate the target output only by adjusting the static linear readout weights.

\begin{figure*}[htbp]
\centerline{\includegraphics[width=6.5in,bb=14 311 568 686]{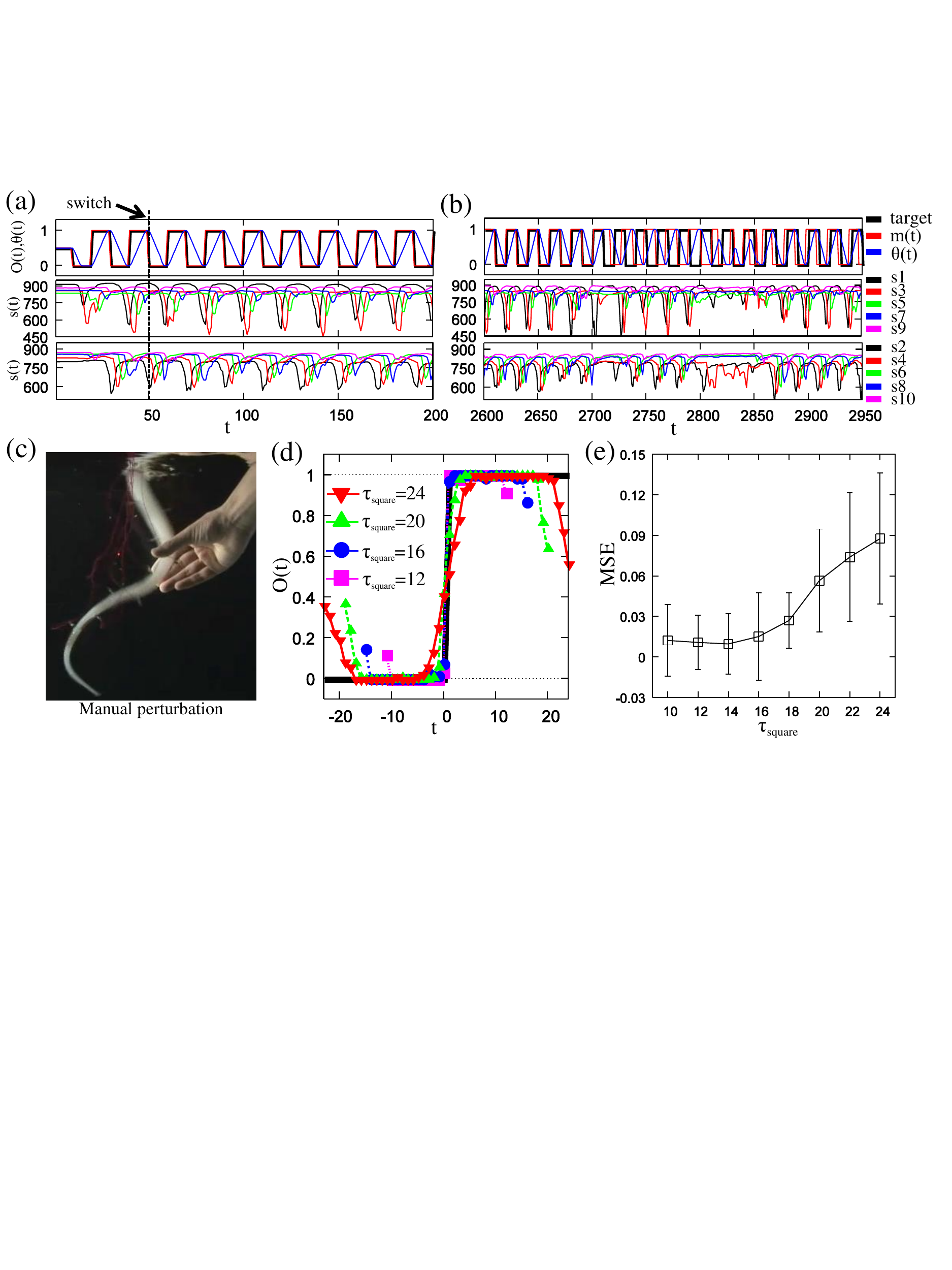}}
\caption{Performance for the closed-loop control task. (a) Plot showing an example of the sensorimotor time series when the system is driven by the closed-loop control with a square function of $\tau_{square} = 10$. The system is initially driven by the target output until $t=50$, and then the loop is closed. The upper diagram plots the time series of the target output, the system output, which is the motor command $m(t)$ in this task, and the base rotation $\theta(t)$. The middle and lower diagrams plot the corresponding sensory time series $s(t)$ in odd and even numbers, respectively. (b) Plot showing an example of the sensorimotor time series when the system is driven by the closed-loop control with the external perturbations in the same experimental condition with (A). The perturbation to the arm is provided at around timestep 2700 to 2850. (c) The external perturbation is provided by the manual mechanical disturbance to the arm. (d) The average system outputs for a single period of a square function driven with open loop providing the target output as input. The plots for $\tau_{square} = 12, 16, 20,$ and $24$ are overlaid. The black line shows the target values as a reference. The time is shifted to set the switching point to $t=0$. (e) The average error (MSE) plot with respect to each $\tau_{square}$. The error bars show the standard deviations. For (d) and (e), the average system output and MSE are calculated by using 50 cycles of oscillations in each condition. See Method section for details.} \label{closedloop}
\end{figure*}

Figure \ref{closedloop}(a) shows an example of a time series with the motor commands and sensory values when the system is driven by the closed-loop control emulating a square function with $\tau_{square}=10$.
The time series of the motor command exactly overlaps with the target output, showing that the closed-loop control is successfully embedded (see also Video S2).
For real-world applications, it is important to investigate whether the system is robust against external perturbation.
We investigated the robustness of the system by applying a manual mechanical perturbation disturbing the arm motion (Fig.\ref{closedloop}(b), (c), and Video S3). 
We found that during the perturbation both the sensory time series and the system output were affected; however, after removing the disturbance, the system was able to recover immediately its original trajectory (Video S3). 
This can be confirmed by checking the time series of the motor commands and their corresponding sensory values, and it implies that our system is robust against external perturbations (Fig.\ref{closedloop}(b)). 
Note that, although the system output shows a phase shift compared with the target output after the perturbation, it is generating a square function with a required length of $\tau_{square}$.

To evaluate the maximal length of $\tau_{square}$ of a square function that our system can embed, we investigated an average system output for one period of a square function by clamping the feedback loop from the system output and providing the
target output as input for each $\tau_{square}$ (Fig.\ref{closedloop}(d), see Method section for details).
If the system is driven by the closed-loop control, the error in the system output would propagate to the motor command through the feedback loop, which makes it difficult to evaluate the limitation of the system performance efficiently. 
In Fig.\ref{closedloop}(e), according to the increase of $\tau_{square}$, the average system output starts to deviate largely from the target output.
By calculating the system error by means of the MSE in this setting, we found that the error grows immediately when $\tau_{square}$ becomes larger than 18 (Fig.\ref{closedloop}(e)).
Consistent with this result, we observed that when $\tau_{square}$ is more than 18, the system cannot embed a correct square function anymore, or it simply stops, continuously providing 0 or 1 as output.
Thus, we can speculate that our system possesses enough memory to be exploited for embedding a square function up to a length of around $\tau_{square} =18$.

\subsection*{Function emulation tasks}
In this final task, we aim to quantitatively characterize the intrinsic computational capacity of our system, particularly focusing on its memory capacity.
By providing a random binary sequence to the motor command as input, the system should perform function emulation tasks using the resulting sensory time series.
Because our system is not an abstract computational unit but has physical and mechanical constraints, we need to define a certain duration of time for one input state or symbol.
We call this duration of time $\tau_{state}$.
We found that when a random binary sequence is provided as motor commands in the form of $\tau_{state} < 5$, the motor overheats and stops.
Accordingly, we performed our experiments with $\tau_{state} \geq 5$.
In addition, we introduced a different time scale for I/O, defined as $t'$, which takes one input symbol as a unit.
This means that $t'$ is increased by increments of 1 for each $\tau_{state}$ timestep (see Method section and Fig.\ref{SI_task3}(a) in it for details).

The first function we aim to emulate is one that provides a delayed version of the input, i.e., $I(t'-n)$ ($n=1, 2,...$) (see Method section for details).
This task enables the direct evaluation of whether the system contains memory traces of a past input within the current sensory values, and is frequently used to evaluate the memory capacity of dynamical systems (see, e.g., \cite{Jaeger3,White,Surya}).
For descriptive purposes, we call this the {\it short-term memory task}.
The second function we aim to emulate is the N-bit parity checker.
The output should provide 0 if $\sum_{d=0}^{n}I(t'-d)$ is an even number; otherwise, it should provide 1, with $n=1, 2,...$ (see Method section for details).
Note that it is actually a ``($n+1$)-bit parity checker'' in our case.
According to the definition, the system needs the memory of input symbols to previous $n$ symbols within the system to emulate this function.
In addition, this function is a nonlinear function, which maps the input to a linearly inseparable state \cite{PDP}.
Because we are only adjusting the static linear weights externally, we can evaluate whether the system contains memory and nonlinearity to be exploited.
This task is also common in the evaluation of the computational capacity of dynamical systems (see, e.g., \cite{EdgeofChaos,Snyder}).
Along with the definition of the input symbol, we also need to determine how to define a corresponding sensory time series.
Let us assume that an input symbol was provided at timestep $t (= t' \tau_{state} )$.
As a result, the arm generates corresponding transient dynamics until the next input symbol is provided at timestep $(t' + 1) \tau_{state}$.
We define sensory values at $(t' + 1) \tau_{state}-1$ as corresponding values $s_{i} (t')$ for this input symbol, which is one timestep before the next input symbol is provided (see Method section and Fig.\ref{SI_task3}(a) in it for details).
By providing random binary input sequences to the system over several trials for each parameter $\tau_{state}$ and $n$, we collected the sensory time series used for training.
In the evaluation, both target functions are simultaneously emulated over a previously unseen random input sequence (see Method section and Fig.\ref{SI_task3}(b) in it for details).

\begin{figure*}[htbp]
\centerline{\includegraphics[width=4.0in,bb=124 295 465 726]{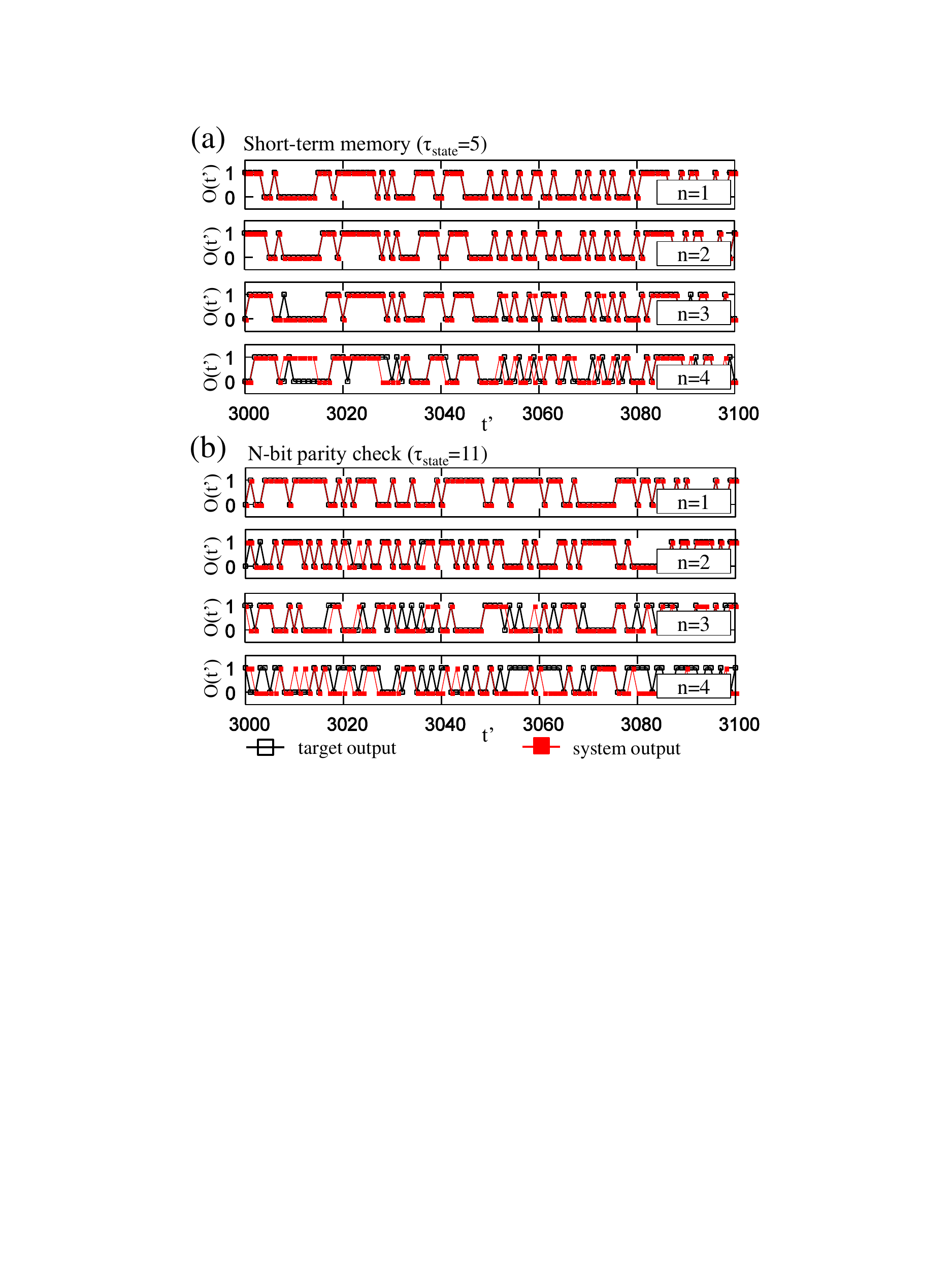}} 
\caption{Examples of the output time series for the function emulation tasks. (a) Plots showing the example of the performance in the short-term memory task with $\tau_{state} = 5$. (b) Plots showing the example of the performance in the N-bit parity check task with $\tau_{state} =11$. For (a) and (b), the black line shows the target outputs and the red line shows the system outputs, and the cases for $n=1, 2, 3,$ and $4$ are shown.} \label{task3_timeseries}
\end{figure*}

Examples of the system performance for the short-term memory task and the N-bit parity check task with $\tau_{state} = 5$ and $11$, respectively, can be found in Fig.\ref{task3_timeseries} and Video S4.
The system output shows almost a perfect match with the target output when $n=1$ and $2$ in $\tau_{state} = 5$ for the short-term memory task (Fig.\ref{task3_timeseries}(a)) and in $\tau_{state} = 11$ for the N-bit parity check task (Fig.\ref{task3_timeseries}(b)).
For both tasks, the performance gradually gets worse when the delay $n$ is increased.
To evaluate the influence of the parameters of $\tau_{state}$ and $n$ on the system performance, we introduced a measure based on mutual information, $MI_{n}$, between the system output and the target output \cite{EdgeofChaos}.
This measure evaluates the similarity between the system output and the target output and, in our experiment, can take the value of 1 as maximum and 0 as minimum.
Additionally, we introduced a measure called ``capacity,'' which is a summation of $MI_{n}$ over the delays, expressed as $C = \sum_{n=1}^{n_{max}}MI_{n}$, where $n_{max}$ is set to 10 in this analysis (see Method section for details).
This measure can evaluate the system's performance over the delays, which can take 10 as maximum and 0 as minimum in our experiment.

\begin{figure*}[htbp]
\centerline{\includegraphics[width=5.0in,bb=78 266 494 737]{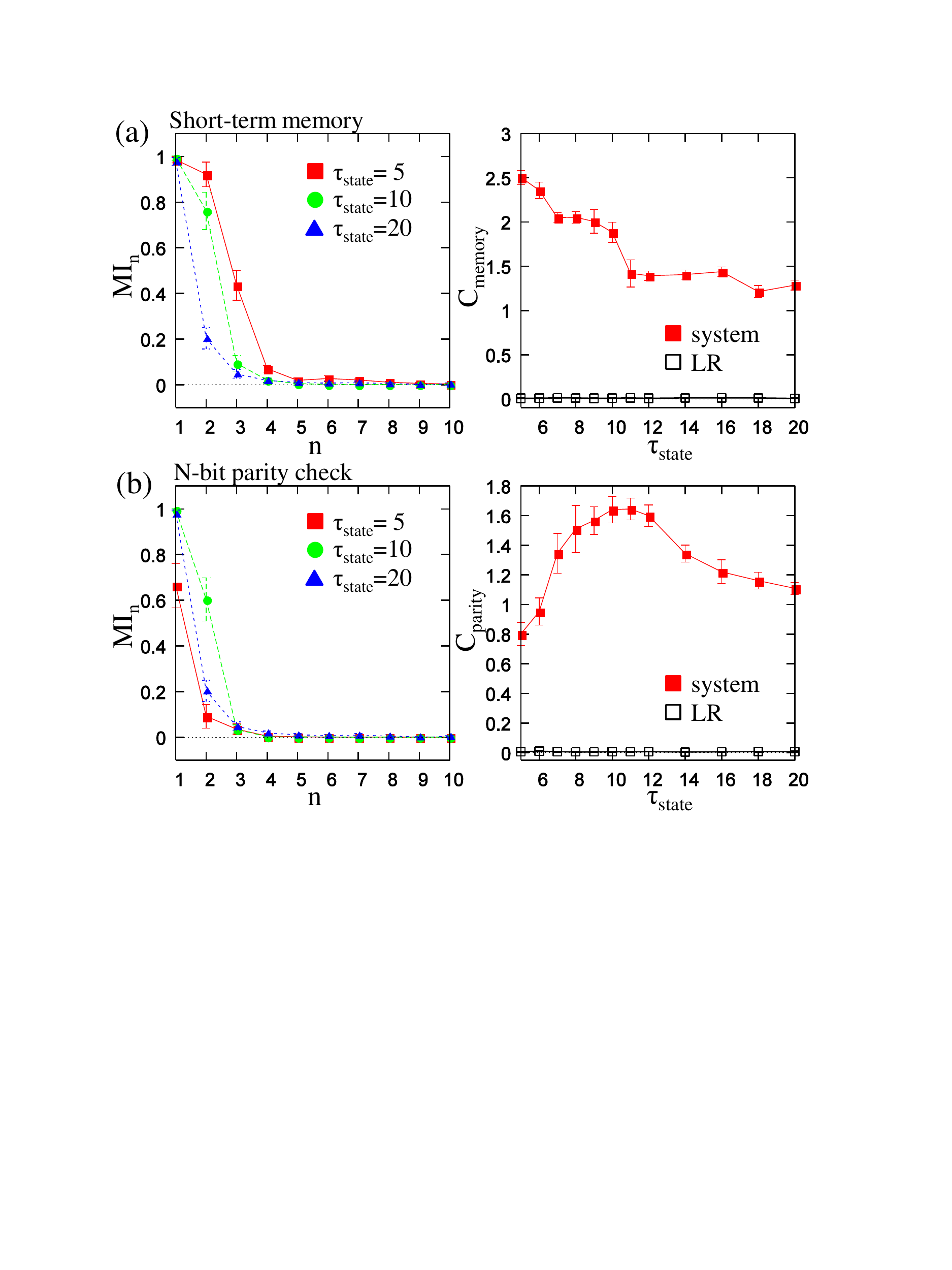}} 
\caption{The average value of $MI_{n}$ according to $n$ (left) and $C$ according to $\tau_{state}$ (right). (a) Plots showing the case for the short-term memory task. (b) Plots showing the case for the N-bit parity check task. Note that the capacities in the short-term memory task and the N-bit parity check task are expressed as $C_{memory}$ and $C_{parity}$, respectively. For each plot on $MI_{n}$, the cases for $\tau=5, 10,$ and $20$ are shown. For each plot on $C$, the results of a logistic regression model (LR) that has a readout directly attached to the input are plotted as comparisons. Results show an almost 0 value for each $\tau_{state}$. The error bars show the standard deviations for each plot.} \label{task3_capacity}
\end{figure*}

Figures \ref{task3_capacity}(a) and (b) show the results of the average $MI_{n}$ for each $n$ value and the average capacity for each $\tau_{state}$ for each task (see Method section for details on the setting).
For the short-term memory task, when $\tau_{state}$ is increased, the value of $MI_{n}$ suddenly drops when $n$ is larger than 2 (Fig.\ref{task3_capacity}(a), left).
For the capacity, increasing $\tau_{state}$ results, first, in a gradual decrease and then in saturation at the constant value for $\tau_{state}>11$ (Fig.\ref{task3_capacity}(a), right).
This can be explained by the behavior of the arm (see Video S4)--if the length of the input symbol is short, it is more likely that the current transient dynamics contains the trace of previously provided input symbols.
Considering that the arm base takes about 9 timesteps to get from one end to the other, if $\tau_{state}$ gets larger than 9 timesteps, the trace of previous input symbols starts to fade out gradually.
Nevertheless, the arm can possess information about the last input symbol because of the simple one-way bend motion.
This explains the maximal performance with respect to $MI_{n}$ when $n=1$.
To see the contribution of the physical body to the computational task, we compared the performance with a model that has a readout directly attached to the input (see Method section for details on the setting).
We can confirm that this model cannot perform this task at all, suggesting that the performance of our system is purely based on the body dynamics (Fig.\ref{task3_capacity}(a), right).

For the N-bit parity check task, even if $\tau_{state}$ is small ($\tau_{state} = 5$), when n = 1 (Fig.\ref{task3_capacity}(b), left), $MI_{n}$ shows a smaller value than when $\tau_{state}$ is larger ($\tau_{state} = 10$ and $20$).
When $\tau_{state}$ gets larger ($\tau_{state} =10$), $MI_{n}$ starts to show the highest value when $n=1$, and a moderately high value in $n=2$.
If we increase $\tau_{state}$ further ($\tau_{state} =20$), $MI_{n}$ still shows the highest value when $n=1$, but the value in $n=2$ starts to decrease.
This tendency reflects the results of the capacity (Fig.\ref{task3_capacity}(b), right).
The capacity shows a peak around $\tau_{state} = 9, 10,$ and $11$.
The low values of capacity in $\tau_{state}$ less than 9 and larger than 11 are because of the low values of $MI_{n}$ in $n=1$ and $n=2$, respectively.
Additionally, in this task, the model with a readout directly attached to the input cannot perform the emulation at all (Fig.\ref{task3_capacity}(b), right).
Considering that the N-bit parity check task requires not only memory but also nonlinearity to perform, this result suggests that, even if the transient dynamics of the arm possesses a high memory capacity when $\tau_{state}$ is low, it does not contain sufficient nonlinearity to be exploited.
This is interesting because this result is not detectable simply by looking at the arm motion.
Furthermore, the results show that the amount of computational capacity depends on the type of motion generated in the arm.

\begin{figure*}[htbp]
\centerline{\includegraphics[width=5.0in, bb=84 353 499 614]{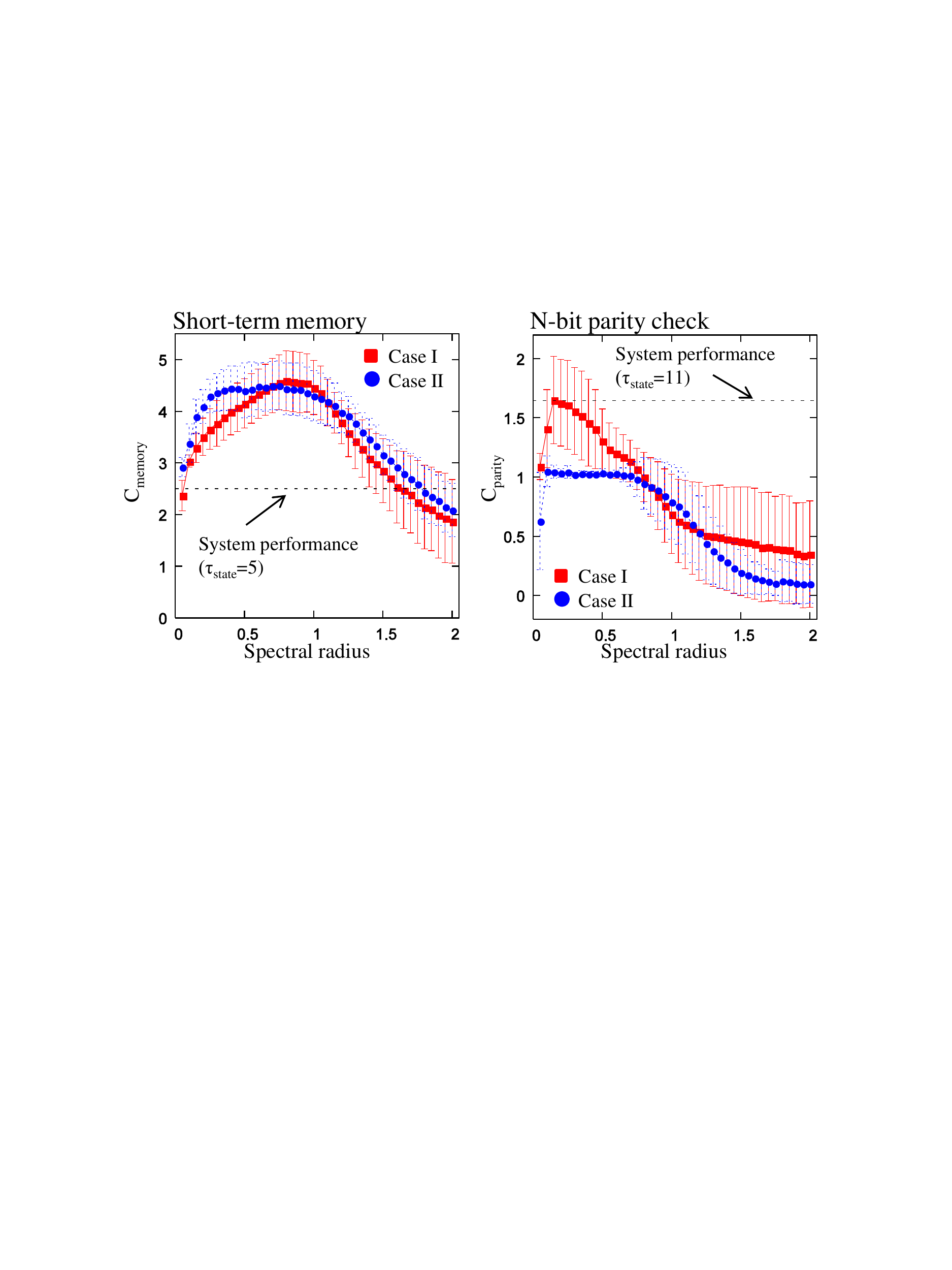}} 
\caption{The average value of $C$ according to the spectral radius in ESN for the short-term memory task (left) and the N-bit parity check task (right). Note that the capacities in the short-term memory task and the N-bit parity check task are expressed as $C_{memory}$ and $C_{parity}$, respectively. 
For each plot, red squares and blue circles show the results for Case I and Case II, respectively, and the highest value of $C$ for our system (when $\tau_{state}=5$ for the short-term memory task and $\tau_{state}=11$ for the N-bit parity check task (Fig.\ref{task3_capacity} in the main text)) is also plotted as a dashed line for comparison.
The error bars show the standard deviations for each plot.} \label{SI_ESN}
\end{figure*}

We have further characterized the computational power of our system by comparing its performance with a conventional ESN, which has the same I/O settings with the same training procedures for the readout weights, the same number of computational nodes (10 fully coupled nodes with one bias term), and the same length of training and evaluation data sets.
It has been shown that the computational performance of an ESN is up to the spectral radius of the reservoir connectivity matrix \cite{Reservoir}.
In each task, we varied the spectral radius of the ESN from 0.05 to 2.0 and calculated the averaged capacity over 30 trials in each spectral radius value, with a new ESN in each trial.
For the short-term memory task, the best capacity value of the ESN was 4.59$\pm$0.58 (Fig.\ref{SI_ESN}, left), while our system showed the best value capacity of 2.50$\pm$0.08 when $\tau_{state}=5$ (Fig.\ref{task3_capacity}(a), right), which was lower than the ESN.
For the N-bit parity check task, the best capacity value of the ESN was 1.65$\pm$0.37 (Fig.\ref{SI_ESN}, right), and the best capacity value of our system showed a similar value of 1.65$\pm$0.07 when $\tau_{state}=11$ (Fig.\ref{task3_capacity}(b), right).
Considering that soft bodies have multifaceted usages and advantages in addition to the computational abilities presented here while the ESN is only focused on computational tasks, we think that our system performance is at a satisfactory level.
Further details on these comparisons are given in Method section.

\section*{Discussion}
In this study, we have systematically demonstrated that the body dynamics of the soft silicone arm can be exploited as computational resources.
In particular, for the closed-loop control task, our results suggest that soft body dynamics can be sufficient to perform the task to control the body without the need of an external controller for additional memory capacity.
This can be, for example, directly applied to the recently proposed octopus-inspired swimming robot \cite{swimming} to generate the arm motion in a closed-loop manner exploiting the body dynamics itself, which largely outsources the computational load required to generate the motor command to the body. 
The technique presented here can be potentially applied to a wide class of soft robots because the main component required is the soft body itself.
Consequently, different types of morphology and material properties of robots that increase the computational capacity of the body should be explored in the future.
In addition, developments in new types of sensors, which can effectively monitor body dynamics, would make the presented approach usable in additional applications.
To conclude, we believe that we have presented a crucial step toward a novel control scheme for soft robots.

In reservoir computing studies, it has been established that to have powerful computational capabilities, a reservoir should have the properties of input separability and fading memory \cite{Maass}. 
Input separability is usually achieved by a nonlinear mapping of the low-dimensional input to a high-dimensional state space. 
Fading memory is a property to uphold the influence of a recent input sequence within the system, which permits integration of stimulus information over time. 
This guarantees reproducible computation, for which the recent history of the signal is important. 
Our insight here was to exploit soft body dynamics as a reservoir.
Passive body dynamics of soft materials typically tend to underactuated systems \cite{underactuated}. 
This naturally maps the actuation signal into the higher dimension of the soft body, which realizes the separability of the actuation signal. 
Furthermore, the interaction between the body and the environment (in our case, the underwater environment) implements fading memory, which takes a certain duration of time to relax when actuated due to the damping effect provided by the environment.
Mechanical structures exhibiting these properties can also be exploited with our approach.

The framework presented in this study may also shed light on the role of the body in biological systems.
Such systems have soft bodies that can adapt and behave effectively in a given ecological niche. 
For example, the octopus does not have any rigid components in its body but it shows extremely sophisticated behavior that capitalizes on its body morphology and muscle structures \cite{Benny}.
In particular, we have shown that a form of short-term memory, which is thought to be a functionality of the brain, can also be found in soft body dynamics.
We think this line of studies is an interesting research direction to be explored further.

\section*{Methods}
\subsection*{Experimental platform setup} 
The experimental platform mainly consists of a soft silicone arm, its actuation, sensing and control systems, and a water tank containing fresh water as the working environment. 
The size of the water tank is 100 cm long, 50 cm wide, and 50 cm deep. 
During experiments, the arm is immersed in the water and actuated by a servo motor at the arm base, which consists of rigid plastic and is directly connected to the motor.
For each experimental trial, the amount of water in the tank is controlled so that it is the same height of the apical surface of the plastic material of the base when the arm is aligned vertically to the water surface. 
Sensors embedded in the arm are used to detect the amount of bending of the arm during experiments. 
The motor commands and sensory data are recorded at each control timestep for further analysis.

\subsection*{A soft silicone arm}
We made the soft arm with silicone rubber (ECOFLEX\texttrademark 00-30 from Smooth-On, Inc.) using an ABS plastic mold manufactured by a 3D printer (Fig.\ref{platform}(a) in the main text and Fig.\ref{arm}(a)). 
The mold has two separate pieces, which can be assembled together. 
The silicone arm has a cone shape and is 44.7 cm long, with a radius of 1.4 cm at one end and a radius of 0.15 cm at the other end (Fig.\ref{arm}(b)), so the arm will not touch the ground and walls during movement.
Ten bend sensors were embedded near the surface of the silicone arm during the process of making the arm. 
Four steps are involved in making a silicone arm with embedded bend sensors: (1) align five bend sensors at the bottom of each piece of the mold; (2) pour a layer of silicone on the bend sensors so that the sensors' arrangement is fixed after the silicone is cured; (3) assemble the two separate pieces of the mold together, and fill the remaining space in the mold with silicone; (4) open the mold and take out the silicone arm after the silicone is cured.

\subsection*{Bend sensors}
To detect the body dynamics of the soft silicone arm, we used flexible, lightweight bend sensors from Flexpoint Sensor Systems, Inc. (Fig.\ref{arm}(c)). 
The size of the sensor is roughly 3.2 cm long, including connectors, 0.7 cm wide, and less than 0.1 cm thick. 
It consists of a thin plastic base film, a layer of coated bend-sensitive ink, and two connectors \cite{sensor1}. 
The coated bend-sensitive ink layer changes its electrical conductivity as the sensor is subjected to bending. 
Therefore, the sensor is actually a potentiometer, which converts mechanical deformation into the change of electric resistance. 
An advantage of the sensor is its large range of resistance change from a few hundred $\Omega$ to a few hundred $K\Omega$; thus, a simple voltage divider can be used to read the sensory output \cite{sensor2}. 
Typical sensor response curves can be found in the design manual of the provider \cite{sensor2}.

\subsection*{Sensory data acquisition system}
We used a sensor board with voltage dividers and a 16-channel multiplexer, an Arduino\texttrademark MEGA 2560 board, and a PC for data acquisition. 
The fixed resistors of the voltage divider are 10 $K\Omega$. 
The multiplexer reads in data from each of its input channels serially and sends it to the Arduino board's analog input pins. 
Then the Arduino board transmits the bend sensors' data to a PC serial port. 
Finally, a Java program running in the PC reads and records the sensory data at each timestep. 
Arranging the electrical cables connecting the sensor connectors and the sensor board is a challenge. 
Because of the repeated bending during experiments, the cables are prone to breakage, especially near the sensor connectors. 
Using an L-shape cable connector and putting the electrical cable completely outside of the arm, the cables are not only easy to change but also become free from bending stress.

\subsection*{Actuation system}
The soft silicone arm is actuated by a Dynamixel RX-64 servo motor, which is controlled by the Java program running on the PC. 
The servo motor is fixed on a plexiglass plate that is placed on top of the water tank. 
The servo motor rotation is transmitted to the soft silicone arm by two identical plastic gears: one is fixed at the end of the motor axle; the other is attached at the larger end of the silicone arm. 
Motor commands used for each experiment are described in the main text.

\subsection*{Experimental procedure for the timer task}
\begin{figure*}[htbp]
\centerline{\includegraphics[width=3.5in, bb=101 70 443 729]{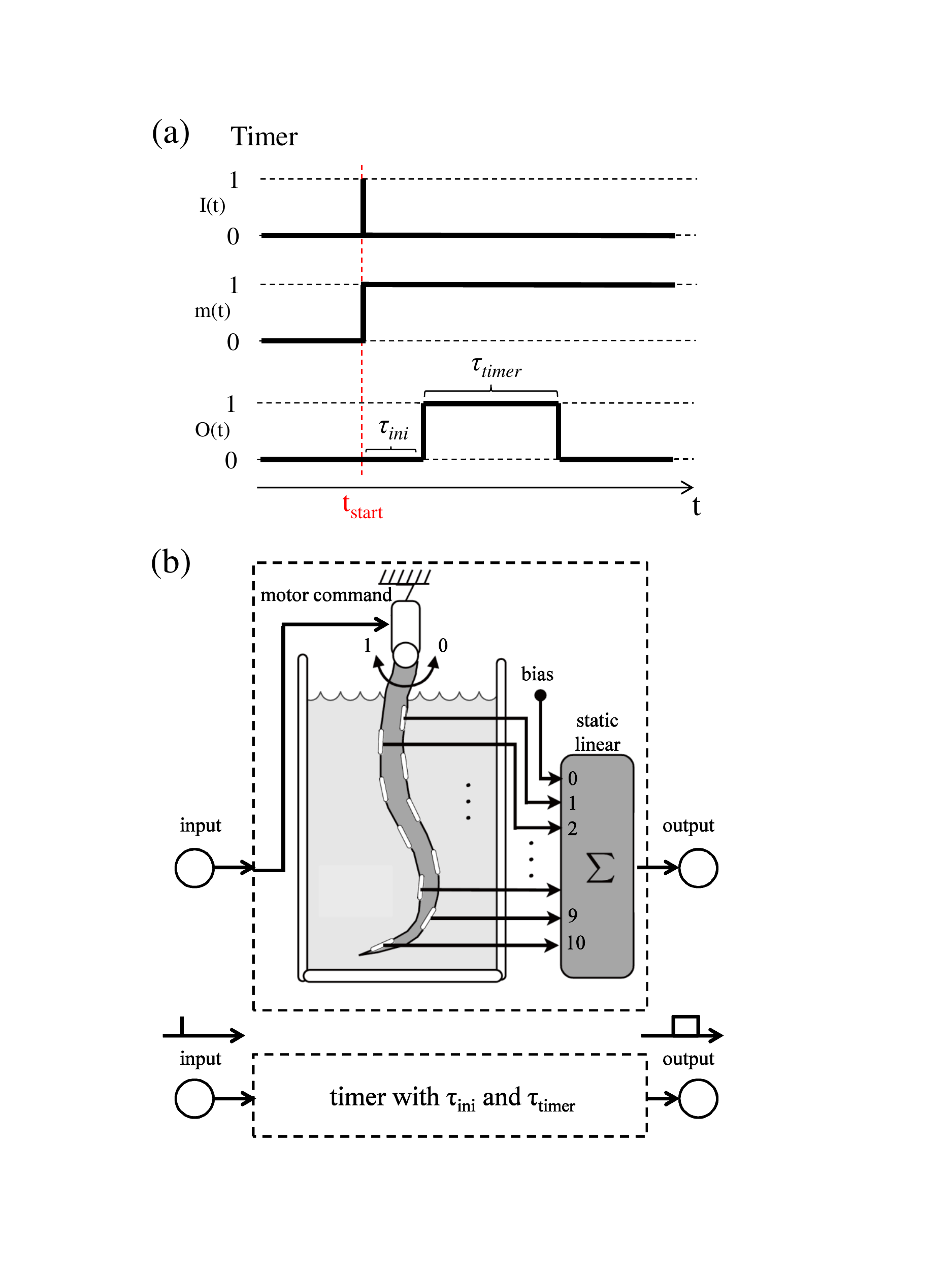}} 
\caption{Experimental procedure for the timer task. (a) Schematics showing the I/O relation with respect to the temporal axis. Triggered by the input at $t_{start}$, the motor command switches from 0 to 1. By exploiting the sensory time series resulting from the soft body dynamics, the system should output a pulse with $\tau_{timer}$ timesteps in length after $\tau_{ini}$ timesteps from timestep $t_{start}$. (b) Schematics showing the generation of the system output. The system output is generated by thresholding the weighted sum of the corresponding sensory time series.} \label{SI_task1}
\end{figure*}

As explained in the main text, our first task was to emulate the function of a timer exploiting the body dynamics of the arm.
The I/O relation for the timer can be expressed as follows (Fig.\ref{SI_task1}(a)):
\begin{align*}
I (t) &= \left \{
\begin{array}{l}
1 \ \ (t=t_{start}) \\
0 \ \ ({\rm otherwise})
\end{array}
\right. \\
O_{target} (t) &= \left \{
\begin{array}{l}
1 \ \ (t_{start}+\tau_{ini} \leq t \leq t_{start}+\tau_{ini}+\tau_{timer}) \\
0 \ \ ({\rm otherwise}).
\end{array}
\right.
\end{align*}
The behavior of the motor commands $m(t)$ according to the input $I(t)$ can be expressed as follows:
\begin{align*}
m (t) &= f(I(t)), \\
&= \left \{
\begin{array}{l}
1 \ \ (t \geq t_{start}) \\
0 \ \ ({\rm otherwise}).
\end{array}
\right.
\end{align*}
Our aim was to emulate this timer by exploiting the body dynamics generated by the input.
Our system outputs are produced by applying static linear readout weights $w_{i}$ ($i=0, 1, ..., 10$) to the sensory time series $s_{i} (t)$ ($i=0, 1, ..., 10$) as follows (Fig.\ref{SI_task1}(b)):
\begin{align*}
O_{system} (t) &= P (\sum_{i=0}^{10} w_{i}s_{i} (t)), \\
P (x) &= \left \{
\begin{array}{l}
1 \ \ (x>0) \\
0 \ \ ({\rm otherwise}),
\end{array}
\right.
\end{align*}
where the system output $O_{system} (t)$ is obtained by thresholding function $P(x)$. 
In this task, the arm is initially set to $L_{right}$, and the motor command is set to 0 ($m(t) = 0$).
At timestep 50, the input provides 1, and triggered by this input command, the motor command switches from 0 to 1 (i.e., $t_{start} = 50$).
After that, the system continues to run for another 200 timesteps.
We consider the overall 250 timesteps as one trial in this task.
For the training procedure, we iterated this process over 25 trials and collected the corresponding sensory time series for each timestep. 
We optimized the linear output weights using these collected sensory time series with a logistic regression to emulate the target output for given $\tau_{ini}$ and $\tau_{timer}$ \cite{Bishop} (see subsequent sections for details).
To evaluate the performance of the system with the optimized weights, we ran 25 additional trials (evaluation trials) and compared the system outputs to the target outputs.
The plot in Fig.\ref{respnse_curve}(c) in the main text is obtained by averaging the system output using the evaluation trials for each timestep.
In Fig.\ref{respnse_curve}(c), the time is shifted so that $t_{start}=0$ for clarity.
In addition, the mean squared error (MSE) in Fig.\ref{respnse_curve}(d) is calculated as follows:
\begin{equation*}
MSE = \frac{1}{T} \sum_{t=0}^{T} (O_{target} (t) - O_{system} (t))^{2},
\end{equation*}
where $T$ is 250 in this task.
For each pair of parameters $(\tau_{ini}, \tau_{timer})$, the readout is trained in the above mentioned manner, and the average MSE is calculated by using the evaluation trials.

\subsection*{Experimental procedure for the closed-loop control task}
In this task, we aimed to embed a square function in a closed loop.
The used square wave function can be expressed as follows (Fig.\ref{SI_task2}(a)):
\begin{equation*}
x (t) = \frac{1}{2}(sgn(\sin(\frac{2\pi}{\tau_{square}} t)) + 1). 
\end{equation*}
To emulate this oscillatory wave pattern by using the sensory time series, the generated output value is fed back as the next motor command to the system and is expressed as follows (Figs.\ref{SI_task2}(b)):
\begin{align*}
m (t) &= I(t), \\
O_{system} (t) &= P (\sum_{i=0}^{10} w_{i}s_{i} (t)), \\
I (t+1) &= O_{system} (t).
\end{align*}
As for the timer task, we emulated the above square wave function only by adjusting the static linear output weights.
Because this task requires feedback to the system, the training procedure is different from the previous one.
During the training phase, we clamped the feedback from the system output, and provided the
target outputs as inputs ($x(t)$ ($= O_{target} (t)$)), which means we set $I(t+1) = x(t)$. 
Thus, the training phase was carried out with an open loop, such that the system was forced into the desired operative state by the target signals in the required $\tau_{square}$ (this approach is typically referred to as {\it teacher forcing}) \cite{Helmut2}. 
In the experimental procedure, the soft silicone arm was first set to the resting state for $\tau_{square}$ timesteps to align the arm vertically to the water surface without motion.
We first ran the system with the teacher forcing condition and collected the corresponding sensory time series data for 2500 timesteps.
The first 100 timesteps were discarded, and the remaining 2400 timesteps were used for training.
After obtaining the optimal readout weights from these collected data using logistic regression, we initialized the system to the resting state and ran the system again with the teacher forcing condition.
After 50 timesteps, we switched the inputs to the system output generated by the trained readout weights and checked whether the system was able to embed the square wave function robustly.

As explained in the main text, to evaluate the performance of the system for each $\tau_{square}$, we ran the system with the teacher forcing condition, collected the sensory time series for 50 cycles of motor oscillations, and trained the weights as previously described. 
Then, by using the optimized weights, we ran a new trial of 50 cycles of motor oscillations as an evaluation phase with the teacher forcing condition and compared the system outputs and the target outputs.
Figure \ref{closedloop}(d) in the main text shows the averaged system output for this evaluation phase.
For the figure, the time is shifted so that the time when the target output switches from 0 to 1 is at $t= 0$ for clarity.
The plot in Fig.\ref{closedloop}(e) in the main text is obtained by averaging the MSE (with $T=2\tau_{square}$) in this evaluation phase.

\subsection*{Experimental procedure for the function emulation tasks}
\begin{figure*}[htbp]
\centerline{\includegraphics[width=3.5in, bb=113 54 444 768]{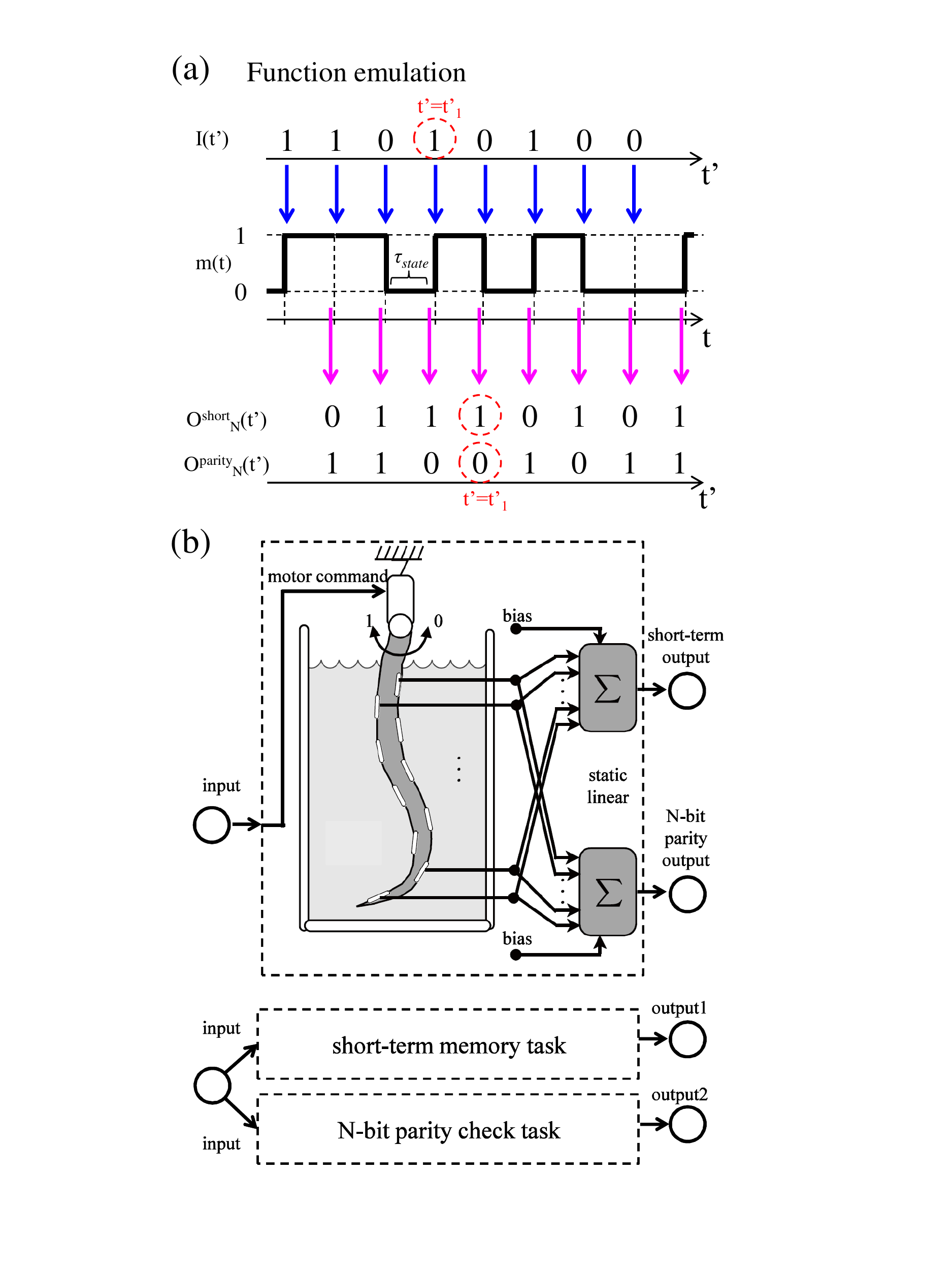}} 
\caption{Experimental procedure for the function emulation tasks. (a) Schematics showing the I/O relation with respect to the temporal axis. The timescale defined for the I/O relation is $t'$. The input symbol is provided to the system for each $\tau_{state}$ timestep. The corresponding sensory time series $s_{i} (t')$ is at timestep $(t' + 1)*\tau_{state} -1$ ($t'_{1}$ is presented as a reference to show the I/O relation). (b) Schematics showing the generation of the system output for function emulation tasks. The system outputs are generated with a multitasking scheme; that is, the same soft body is employed to carry out two different tasks at the same time.} \label{SI_task3}
\end{figure*}

In this task, we aimed to emulate the short-term memory task and the N-bit parity check task. 
Following the notation defined in the main text, the function for the short-term memory task can be expressed as follows (Fig.\ref{SI_task3}(a)):
\begin{equation*}
O^{short}_{n} (t') = I(t' -n),
\end{equation*}
where $I(t')$ is a random binary sequence.
The function for the N-bit parity check task is expressed as follows (Fig.\ref{SI_task3}(a)):
\begin{align*}
O^{parity}_{n} (t') &= Q(\sum_{d=0}^{n}I(t' -d)), \\
Q (x) &= \left \{
\begin{array}{l}
0 \ \ (x \equiv 0(mod 2)) \\
1 \ \ ({\rm otherwise}).
\end{array}
\right.
\end{align*}
Here, the system output is generated as follows (Fig.\ref{SI_task3}(b)):
\begin{equation*}
O_{system} (t') = P(\sum_{i=0}^{10} w_{i}s_{i} (t')).
\end{equation*}
In this task, one trial consists of 3500 symbols (i.e., 3500*$\tau_{state}$ timesteps).
The first 100 symbols are discarded, the next 2400 symbols are used for training, and the last 1000 symbols are used for the system evaluation. 
Both functions are emulated simultaneously using the same random input sequence.
This is typically referred to as {\it multitasking} (Fig.\ref{SI_task3}(b)). 
The training of the static linear readout weights is conducted by a logistic regression for each function emulation task. 
For each $\tau_{state}$, we ran the system for 10 trials and evaluated the system performance with the target output for each $n$ by using the mutual information expressed as follows (the same evaluation scheme is introduced in \cite{EdgeofChaos}):
\begin{align*}
MI_{n} &= \sum_{O_{system}(t) \in O} \sum_{O_{target}(t) \in O} p(O_{system}(t), O_{target}(t)) \\
&\log \frac{p(O_{system}(t), O_{target}(t))}{p(O_{system}(t)) p(O_{target}(t))},
\end{align*}
where $O= \{ 0, 1\} $, and $p(x)$ and $p(x, y)$ are the probability of $x$ and the joint probability of $x$ and $y$, respectively.
The base of $log$ is fixed to 2 throughout this study.
When $O_{system}(t)$ and $O_{target}(t)$ are independent, then $MI_{n} =0$.
When $O_{system}(t)$ is exactly the same as $O_{target}(t)$, then $MI_{n} =1$ because $O_{target}(t)$ is random.
As explained in the main text, we introduce a measure called ``capacity'' $C$, which is a summation of $MI_{n}$ over the delays, expressed as $C = \sum_{n=1}^{n_{max}}MI_{n}$, where $n_{max}$ is set to 10 in this task.
In Fig.\ref{task3_capacity} in the main text, $MI_{n}$ and the capacity $C$ are averaged over 10 trials in each condition.

To see the contribution of the physical body to the computational task, 
we compared the performance with a logistic regression model that has a readout directly attached to the input, expressed as
\begin{equation*}
O_{LR} (t') = P(w_{1}I (t') + w_{0}),
\end{equation*}
where the weights are trained by a logistic regression using the same time series as in the training phase for each function emulation task.
The capacity $C$ for the logistic regression model, which is shown in Fig.\ref{task3_capacity} in the main text, is calculated and averaged over 10 trials in each condition as well.

To further characterize the computational power of our system, we have compared its task performance with a conventional echo state network (ESN) \cite{Jaeger1}.
The ESN is a type of recurrent neural network, which has $N$ internal network units, input units, and output units.
Activation of the $i$th internal unit is $x_{i}(t')$ ($i=1, ..., N$), and activation of the input and output units are $I_{ESN}(t')$ and $O_{ESN}(t')$, respectively. 
Throughout this analysis, $N$ is set to 10, which is the same number of sensors in our system.
We used the same I/O setting for the ESN as with our system, which takes a binary state, to compare the task performance directly.
For the random binary input sequence, we adopted two cases. 
In the first case (Case I), we directly projected the $\{ 0, 1 \}$- binary state input $I(t')$ to the internal network units, which can be denoted as $I_{ESN}(t') = I(t')$.
For the second case (Case II), we changed the actual input value to $I_{ESN}(t') = -1$ only if $I(t')=0$; otherwise, $I_{ESN}(t') = 1$. 
In the ESN, if $I_{ESN}(t') = 0$, the internal units receive no external input, and therefore, are expected to introduce an asymmetry into the network performance.
To avoid the bias introduced by this asymmetric nature of input $I(t')$ in our comparison, we adopted these two cases.
For both cases, connection weights for the $N \times N$ internal network connecting $i$th unit with $j$th unit are denoted as $w_{ij}$, and connection weights going from the input unit into the $i$th internal unit are denoted as $w^{in}_{i}$.
The readout weights $w^{out}_{i}$ go from $N$ internal units and 1 bias to the output unit (where $w^{out}_{0}$ is a bias term).
The output weights $w^{out}_{i}$ are trained in the same procedure explained above for each task, and the internal weights $w_{ij}$ and the input weights $w^{in}_{i}$ are randomly assigned from the range $[-1.0, 1.0]$ and fixed beforehand.
The activation of internal units and the output unit are updated as
\begin{align*}
x_{i} (t') &= f (\sum_{j=1}^{10} w_{ij}x_{j}(t'-1) + w^{in}_{i}I_{ESN}(t')), \\
O_{ESN}(t') &= P(\sum_{i=1}^{10} w^{out}_{i}x_{i} (t') + w^{out}_{0}),
\end{align*}
where $f$ is a $\tanh$ function.
As explained previously, one trial consists of 3500 symbols, and the first 100 symbols are discarded, the next 2400 symbols are used for training, and the last 1000 symbols are used for the ESN evaluation.
In the training phase, a white noise in the range of $[-10^{-4}, 10^{-4}]$ is added to the internal states.  
The weights are trained by the logistic regression, and the multitasking scheme is adopted as well, where both functions are emulated simultaneously using the same random input sequence.
It is reported that the computational power of ESN can be well characterized by the spectral radius of the internal connection weight matrix \cite{Reservoir}.
In this experiment, we varied the spectral radius from 0.05 to 2 and observed the ESN performance in terms of the capacity.
The capacity is averaged over 30 trials with a new ESN each (renewing the internal weights $w_{ij}$ and the input weights $w^{in}_{i}$) for each setting of the spectral radius.

Figure \ref{SI_ESN} shows the average capacity of ESN according to the spectral radius for each task.
In the short-term memory task, we can see that the performance of ESN is relatively much better than our system performance (the performance of ESN becomes worse than our system only when the spectral radius is 0.05 or larger than 1.65 in Case I, and when larger than 1.75 in Case II), while in the N-bit parity check task, the performance of ESN is much worse than our system performance (the performance of ESN becomes better than our system only when the spectral radius is 0.15 in Case I).

\subsection*{Training the readout weights using logistic regression}
In this section, we provide a brief overview of logistic regression and how it is used to train the readouts from the sensor to produce the desired output.
Detailed information on the logistic regression can be found in \cite{Bishop}.
As described in the main text, our system output takes binary states for each task by thresholding the value obtained by summing up the linearly weighted sensory values.
This is basically a two-class classification of the sensory values, which can be appropriately dealt with using the logistic regression model.
Following the notation used in the main text, we have 11 sensory time series $s_{i}(t)$ $(i=0, 1, ..., 10)$ and corresponding 11 linear and static readout weights $w_{i}$ $(i=0, 1, ..., 10)$.
Note that, as explained in the main text, a unit expressing time for the function emulation tasks was $t'$ due to an input symbol introduced to have a specific duration of time, that is, $\tau_{state}$. 
In the following descriptions, we keep using $t$ for the general case, but one can replace $t$ with $t'$ for the function emulation tasks.
We also introduce the vectors ${\bf s_{t}} = [s_{0} (t), s_{1} (t), ..., s_{10} (t)]^{\mathrm{T}}$ and ${\bf w} = [w_{0}, w_{1}, ..., w_{10}]^{\mathrm{T}}$ for descriptive purposes. Our aim here is to optimize ${\bf w}$ by using corresponding $T_{train}$ data pairs, ${\bf s_{t}}$ and $O_{target} (t)$, collected in the training phase in each task. For example, in the first task of constructing a timer, we had 25 trials, which had 250 timesteps each, as training data. This results in $T_{train} = 250 \times 25 = 6250$ data pairs. In the closed-loop control task and the function emulation tasks, $T_{train} = 2400$ data pairs were used for training (see previous sections).

Based on \cite{Bishop}, we start by introducing a posterior probability for classes $C_{1}$ and $C_{2}$; these classes correspond to the output states 1 and 0, respectively,
\begin{align*}
p(C_{1} | {\bf s_{t}}) &= y({\bf s_{t}}) = \sigma ({\bf w}^{\mathrm{T}}{\bf s_{t}}), \\
p(C_{2} | {\bf s_{t}}) &= 1 - p(C_{1} | {\bf s_{t}}),
\end{align*}
where $\sigma (\cdot)$ is the logistic sigmoid function. We now make use of the maximum likelihood method to determine the optimal weights.
For a data set $({\bf s_{t}}, O_{target} (t))$, where $O_{target} (t) \in \{ 0, 1 \}$ and $t=1, ..., T_{train}$, the likelihood function can be expressed as
\begin{align*}
p( {\bf O_{target}} | {\bf w}) = \prod_{t=1}^{T_{train}} y_{t}^{O_{target} (t)} (1-y_{t})^{1-O_{target} (t)},
\end{align*}
where ${\bf O_{target}} = [O_{target} (1), O_{target} (2), ..., O_{target} (T_{train})]^{\mathrm{T}}$ and $y_{t} = p(C_{1} | {\bf s_{t}})$.
The error function of this likelihood can be expressed as
\begin{align*}
E({\bf w}) &= -\ln{p( {\bf O_{target}} | {\bf w})} \\
&= -\sum_{t=1}^{T_{train}} \{ O_{target} (t) \ln{y_{t}} + (1-O_{target} (t)) \ln{(1-y_{t})} \},
\end{align*}
where $y_{t} = \sigma (a_{t})$ and $a_{t} = {\bf w}^{\mathrm{T}}{\bf s_{t}}$. Taking the gradient and Hessian of this error function, we obtain
\begin{align*}
\nabla E({\bf w}) &= \sum_{t=1}^{T_{train}} (y_{t} - O_{target} (t)){\bf s_{t}} = {\bf S}^{\mathrm{T}} ({\bf y} - {\bf O_{target}}), \\
{\bf H} &= \nabla \nabla E({\bf w}) = \sum_{t=1}^{T_{train}} y_{t}(1-y_{t}){\bf s_{t}}{\bf s_{t}}^{\mathrm{T}} = {\bf S}^{\mathrm{T}} {\bf R} {\bf S},
\end{align*}
where ${\bf y} = [y_{1}, y_{2}, ..., y_{T_{train}}]^{\mathrm{T}}$ and ${\bf S}$ is a matrix expressed as
\begin{eqnarray*}
{\bf S} = \left[
\begin{array}{c}
{\bf s_{1}}^{\mathrm{T}} \\
\vdots \\
{\bf s_{T_{train}}}^{\mathrm{T}} \\
\end{array}
\right] = \left[
\begin{array}{ccc}
s_{0} (1) & \cdots & s_{10} (1) \\
\vdots & \ddots & \vdots \\
s_{0} (T_{train}) & \cdots & s_{10} (T_{train}) \\
\end{array}
\right].
\end{eqnarray*}
Additionally, we introduce the $T_{train} \times T_{train}$ diagonal matrix ${\bf R}$ with elements, $R_{tt} = y_{t} (1-y_{t})$.
To minimize the function $E({\bf w})$, we used the iterative reweighted least squares method. The Newton-Raphson update formula for the logistic regression model is expressed as
\begin{align*}
{\bf w}^{(new)} &= {\bf w}^{(old)} - {\bf H}^{-1} \nabla E({\bf w}) \\
&= {\bf w}^{(old)} - ({\bf S}^{\mathrm{T}} {\bf R} {\bf S})^{-1}{\bf S}^{\mathrm{T}} ({\bf y} - {\bf O_{target}}) \\
&= ({\bf S}^{\mathrm{T}} {\bf R} {\bf S})^{-1} \{ {\bf S}^{\mathrm{T}} {\bf R} {\bf S}{\bf w}^{(old)} - {\bf S}^{\mathrm{T}} ({\bf y} - {\bf O_{target}}) \} \\
&= ({\bf S}^{\mathrm{T}} {\bf R} {\bf S})^{-1} {\bf S}^{\mathrm{T}} {\bf R} {\bf z},
\end{align*}
where ${\bf z}$ is a $T_{train}$ -dimensional vector with elements
\begin{align*}
{\bf z} = {\bf S} {\bf w}^{(old)} - {\bf R}^{-1} ({\bf y} - {\bf O_{target}}).
\end{align*}
We apply this procedure iteratively, each time using the new weight vector ${\bf w}$ to compute an updated weighing matrix ${\bf R}$ until $\frac{ \|{\bf w}^{(new)} -{\bf w}^{(old)} \| }{\| {\bf w}^{(old)} \|}$ is less than 0.5 in all our experiments.

\begin{acknowledgments}
This work was partially supported by the European Commission in the ICT-FET OCTOPUS Integrating Project (EU project FP7-231608), and by the JSPS Postdoctoral Fellowships for Research Abroad.
\end{acknowledgments}

\end{document}